\documentclass[sigconf]{acmart}

\AtBeginDocument{%
	\providecommand\BibTeX{{%
			\normalfont B\kern-0.5em{\scshape i\kern-0.25em b}\kern-0.8em\TeX}}}

\setcopyright{acmcopyright}
\copyrightyear{2020}
\acmYear{2020}
\acmDOI{XX.XXXX/XXXXXX.XXXXXX}

\acmConference[SIGIR '20]{Xi'an '20: Proceedings of the 43rd International ACM SIGIR Conference on Research and Development in Information Retrieval}{July 25--30, 2020}{Xi'an, China}
\acmBooktitle{Proceedings of the 43rd International ACM SIGIR Conference on Research and Development in Information Retrieval (SIGIR '20), July 25--30, 2020, Xi'an, China}
\acmPrice{15.00}
\acmISBN{978-1-4503-XXXX-X/18/06}

\usepackage{tabularx,ragged2e}
\usepackage{booktabs}
\usepackage[normalem]{ulem}
\usepackage{multirow}
\usepackage{arydshln}

\newcommand{\beforetable}{\vspace{-10pt}}
\newcommand{\aftertable}{\vspace{-10pt}}

\newcommand{\zla}{\textsc{Dexa}}

\newcommand{\tra}{\textsc{Control}}
\newcommand{\crowd}{\textsc{Control}}

\begin{document}

\title[DEXA: Supporting Non-Expert Annotators with  Dynamic Examples from Experts]{DEXA: Supporting Non-Expert Annotators \\ with  Dynamic Examples from Experts}

\author{Markus Zlabinger, Marta Sabou, Sebastian Hofst{\"a}tter, Mete Sertkan, and Allan Hanbury}
\email{first.last@tuwien.ac.at}
\affiliation{\institution{TU Wien, Vienna, Austria}}

%
%
%

\renewcommand{\shortauthors}{}

\settopmatter{printacmref=true}
\fancyhead{}

\begin{abstract}
The success of crowdsourcing based annotation of text corpora depends on ensuring that crowdworkers are sufficiently well-trained to perform the annotation task accurately. To that end, a frequent approach to train annotators is to provide instructions and a few example cases that demonstrate how the task should be performed (referred to as the \tra{} approach). These globally defined \textit{task-level examples}, however, (i) often only cover the common cases that are encountered during an annotation task; and (ii) require effort from crowdworkers during the annotation process to find the most relevant example for the currently annotated sample. To overcome these limitations, we propose to support workers in addition to task-level examples, also with \textit{task-instance level} examples that are semantically similar to the currently annotated data sample (referred to as Dynamic Examples for Annotation, \zla{}). Such dynamic examples can be retrieved from collections previously labeled by experts, which are usually available as gold standard dataset. We evaluate \zla{} on a complex task of annotating participants, interventions, and outcomes (known as PIO) in sentences of medical studies. The dynamic examples are retrieved using BioSent2Vec, an unsupervised semantic sentence similarity method specific to the biomedical domain. Results show that (i) workers of the \zla{} approach reach on average much higher agreements (Cohen's Kappa) to experts than workers of the the \tra{} approach (avg. of 0.68 to experts in \zla{} vs. 0.40 in \tra{}); (ii) already three per majority voting aggregated annotations of the \zla{} approach reach substantial agreements to experts of 0.78/0.75/0.69 for P/I/O (in \tra{} 0.73/0.58/0.46). Finally, (iii) we acquire explicit feedback from workers and show that in the majority of cases (avg. 72\%) workers find the dynamic examples useful.
\end{abstract}

\ccsdesc[500]{Information systems~Crowdsourcing}
\ccsdesc[500]{Information systems~Test collections}

\keywords{human data annotation, crowdsourcing, PICO task}

\maketitle

\newcommand{\sectionspace}{\vspace{-8pt}}

\sectionspace{}
\section{Introduction}
The success of crowdsourcing based annotation of text corpora depends on ensuring that crowdworkers are sufficiently well-trained to perform the annotation task accurately. Reaching a certain quality threshold is challenging, especially in tasks that require specific expertise to be performed (e.g. in the medical domain~\cite{nyeCorpusMultiLevelAnnotations2018}).

The common approach to compensate the missing knowledge of individual non-expert workers is to train them via task instructions and a few example cases that demonstrate how the task should be performed~\cite{nyeCorpusMultiLevelAnnotations2018,snowCheapFastIt2008} (referred to as the \tra{} approach). These globally defined \textit{task-level} examples, however, often (i) only cover the common cases that are encountered during an annotation task and (ii) require effort from crowdworkers during the annotation process to find the most relevant example for the currently annotated sample.

In this paper, we address these limitations with a new annotation approach called \textbf{D}ynamic \textbf{Ex}amples for \textbf{A}nnotation (\zla{}). In addition to task-level examples, annotators are supported with \textit{task-instance level} examples that are semantically similar to the currently annotated sample. The task-instance examples are retrieved from data samples previously annotated by experts. Such expert samples are usually available since they are crucial to measure the quality of non-expert annotators~\cite{snowCheapFastIt2008,danielQualityControlCrowdsourcing2018,doroudiLearningScienceComplex2016}. We propose to split the expert samples into training samples from which dynamic examples are retrieved and test samples which are injected into the annotation process to measure worker performance.

We apply the \zla{} approach on a task of the medical domain, known as the PIO\footnote{The difference to the PICO task is that Intervention/Control are not differentiated~\cite{nyeCorpusMultiLevelAnnotations2018}} task -- where annotators label the Participants (P), Interventions (I), and Outcomes (O) in clinical trial reports. Specifically, we ask non-expert annotators to highlight the exact text phrases that describe either\footnote{To reduce overhead for workers, we split the PIO task into 3 individual sub-tasks.} P, I, or O within the sentences of clinical trial reports. The trial reports used in our experiments stem from the EBM-Corpus~\cite{nyeCorpusMultiLevelAnnotations2018}, for which gold standard PIO labels are available. For the retrieval of dynamic examples, we use BioSent2Vec~\cite{chenBioSentVecCreatingSentence2019}, an unsupervised semantic short-text similarity method specific to the biomedical domain.

We compare \zla{} to the \tra{} approach with respect to the annotation quality of individual workers and the annotation quality of aggregated (e.g. majority vote) redundant annotations from multiple workers. To measure the annotation quality of non-expert workers, we compute the inter-annotator agreement to the gold standard labels using Cohen's Kappa. Our results show that (i) workers using the \zla{} approach reach on average $0.28$ higher agreements to experts than workers using the \tra{} approach (avg. of 0.68 in \zla{} vs. 0.40 in \tra{}); (ii) three per majority voting aggregated annotations of the \zla{} approach already lead to substantial agreements to experts of 0.78/0.75/0.69 for P/I/O (in \tra{} 0.73/0.58/0.46). Finally, (iii) we acquire explicit feedback from workers on the usefulness of the dynamic examples and show that in the majority of cases (avg. 72\%) workers find the dynamic examples useful. For these useful examples, they reach a higher agreement to experts of $0.24$ (avg. over all PIO) than for other examples.

\noindent The contributions of this paper are:
\begin{itemize}
	\item We propose \zla{}, a new annotation approach for the collection of high-quality annotations from non-experts.
	\item We apply the approach to the complex PIO annotation task and show high agreements between non-experts and experts.
	\item We make the collected crowdsourcing annotations and the code used for experiments available at \url{https://github.com/Markus-Zlabinger/pico-annotation}
\end{itemize}
After discussing related work (Sec.~\ref{sec:related}), we describe the DEXA approach (Sec.~\ref{sec:dexa}) and its evaluation on the PIO task (Sec.~\ref{sec:pio} and~\ref{sec:results}).

\sectionspace{}
\section{Related Work}
\label{sec:related}
A common strategy to obtain higher quality labels from non-expert annotators is to redundantly collect annotations for each data sample, and then apply an aggregation method to create a final label that is of a higher quality than the individual labels~\cite{shengGetAnotherLabel2008,snowCheapFastIt2008,sabouCrowdsourcingResearchOpportunities2012}. A simple aggregation method is to conduct a majority voting. More sophisticated methods aim to identify reliable annotators and weight their annotations as more important than the annotations of less reliable annotators~\cite{dawidMaximumLikelihoodEstimation1979}. Note that the \zla{} approach can be combined with aggregation strategies, as we do in our experiments.

To improve the quality of individual annotators, several techniques are summarized by Daniel et al.~\cite{danielQualityControlCrowdsourcing2018}. For example, improving the worker motivation (e.g. higher payment), task simplification, providing constant feedback to workers, or filtering of unreliable workers. Besides these techniques, various annotation approaches are proposed.  For example, Kobayashi et al.~\cite{kobayashiEmpiricalStudyShortand2018} allow workers to change their annotation of a data sample after showing how other workers have annotated the sample. By examining the samples of other crowdworkers, a learning effect is induced in the crowdworkers increasing their accuracy for the annotation of future samples. Suzuki et al.~\cite{suzukiAtelierRepurposingExpert2016} propose a system where inexperienced annotators can seek advice from experts, so-called mentors. Through mentoring, inexperienced annotators should obtain the skills that are required for a task and produce labels of high quality.

While the presented literature studies various aspects of improving the quality of individual non-expert annotators, little is known about how to effectively present demonstration examples~\cite{doroudiLearningScienceComplex2016} and whether such samples are effective in increasing the annotation quality. We give new insights into this topic in this study.

\sectionspace{}
\section{Dynamic Examples for Annotation}
\label{sec:dexa}
In this section, we describe our novel annotation approach, called \textbf{D}ynamic \textbf{Ex}amples for \textbf{A}nnotation (\zla{}). We show examples to annotators on a task-instance-level, i.e., dynamic to the current sample instance that is annotated. Given a set of labeled expert samples $E$ and a set of samples $U$ to be labeled by non-experts, the \zla{} annotation approach consists of following steps: 

\begin{enumerate}
	\item The samples of $E$ are divided into a test set $E_{te}\subset E$ and a training set $E_{tr} \subset E$, where $E_{te} \cap E_{tr} = \emptyset$. From the training set, the dynamic examples are drawn. The samples from the test set are injected into $U$ to measure the quality of the non-expert annotators, resulting in the annotation set $A=U \cup E_{te}$. 
	\item An unsupervised similarity method $\textrm{sim}(p,a) \in \mathbb{R}$ is selected to compute the semantic similarity between a sample $p \in E_{tr}$ of the training set to a sample $a \in A$ of the annotation set. The similarity method should be selected based on the task at hand. For example, in our experiments, samples are sentences, and therefore, we use a semantic sentence-to-sentence similarity method, as described in Section~\ref{sec:piodexa}.
	\item The annotation set $A$ is labeled by non-experts. For each unlabeled sample $a$, the similarity method $\textrm{sim}(p,a)$ is used to compute the similarity to each sample in the training set, i.e, $\textrm{sim}(p_1,a), \ldots, \textrm{sim}(p_n,a) \forall p \in E_{tr}$. Then, the top $k$ most similar samples $p_1,\ldots, p_k \in E_{tr}$ are shown as dynamic demonstration examples to the annotators.
	\item Finally, the accuracy of non-expert annotators is compared to that of expert annotators based on the test samples $E_{te}$ that were injected into the annotation set $A$ in step (1).
\end{enumerate}

\section{Evaluating DEXA on PIO tasks}
\label{sec:pio}
In the PIO annotation task~\cite{huangEvaluationPICOKnowledge2006,nyeCorpusMultiLevelAnnotations2018}, annotations are collected for the \textbf{P}articipants (e.g., "patients with headache"), \textbf{I}nterventions (e.g., "ibuprofen"), and \textbf{O}utcomes (e.g., "pain reduction") of medical studies. Due to the complexity of this task, PIO annotations were initially only annotated on a binary sentence level~\cite{kimAutomaticClassificationSentences2011}, where a sentence was labeled whether it contained a P, I, or O. Recently, fine-grained text span annotations were collected~\cite{nyeCorpusMultiLevelAnnotations2018}, with annotators highlighting the exact text phrases within a sentence that describe P, I, or O. However, using standard task-level training for this task resulted in non-expert workers reaching only weak agreements compared to experts~\cite{nyeCorpusMultiLevelAnnotations2018}. To evaluate \zla{}, we apply it in the setting of~\cite{nyeCorpusMultiLevelAnnotations2018} where we augment task-level examples with dynamic task-instance level examples.

\subsection{Dataset}
We consider the 191 clinical trial reports of the EBM-NLP corpus~\cite{nyeCorpusMultiLevelAnnotations2018}, where for each trial and PIO element gold standard labels are available. The reports originate from PubMed and consist of a title and an abstract. As preprocessing steps, we use the Stanford CoreNLP~\cite{manningStanfordCoreNLPNatural2014} to segment and NLTK~\cite{birdNaturalLanguageProcessing2009} to tokenize the sentences. Next, we split the 191 reports into test set $E_{te}$ for evaluation (41 reports with 426 sentences) and training set $E_{tr}$ (150 reports with 1,636 sentences), from which dynamic examples are retrieved for the \zla{} approach. Note that the test sentences are usually injected into a much larger set $U$ for which no gold labels are available (see Step (1) in Section \ref{sec:dexa}); however, in this study, we aim to evaluate our annotation approach and therefore only sentences are annotated that overlap with the gold standard.

\subsection{Annotation Setup}
\label{sec:piosetup}
We follow the annotation setup described in~\cite{nyeCorpusMultiLevelAnnotations2018} with crowdworkers hired from the Amazon Mechanical Turk (AMT). Annotations for P, I and O are divided into three individual sub-tasks to reduce the cognitive overhead for workers. For each sub-task, annotation instructions and a few task-level examples are provided to workers, available as an appendix in~\cite{nyeCorpusMultiLevelAnnotations2018}. Workers are allowed to participate in one of the sub-tasks if their work approval rate for previous tasks is at least 90\%. A small-scale test run is performed to filter out spammers and workers who do not follow the task instructions. Workers who pass the test run qualify for the full-scale run.

\subsection{\zla{} Approach}
\label{sec:piodexa}
Within the annotation setup described above, we apply the \zla{} approach to collect non-expert labels for the 426 test sentences. We develop an annotation interface that can be embedded as a design layout in the AMT platform. In each HIT, we ask workers to annotate, depending on the sub-task, either P, I, or O within a sentence. For each sentence, we present three dynamic examples ($k=3$), and we acquire feedback from workers on whether they found at least one of these examples useful to support their annotation work.

The dynamic examples that we show to support annotators are retrieved from the training set using the sentence embedding model BioSent2Vec~\cite{chenBioSentVecCreatingSentence2019}. Specifically, we compute the cosine similarity $\textrm{sim}(p,a)=\textrm{cos}(\mathbf{e}_p,\mathbf{e}_a) \in \mathbb{R}$, where $\mathbf{e}$ refers to the BioSent2Vec embedding of the sentences $p$ and $a$. We use BioSent2Vec since (i) it is the state-of-the-art for various short-text similarity tasks in the biomedical domain, and (ii) a pre-trained model is available\footnote{\url{https://github.com/ncbi-nlp/BioSentVec}} trained on PubMed~\cite{chenBioSentVecCreatingSentence2019}, which is the same underlying data source as the clinical trial reports used in our study.

\newcommand{\ielem}[1]{{\color[RGB]{0,128,0}\uuline{#1}}}
\newcommand{\pelem}[1]{{\color[RGB]{200,0,0}\uline{#1}}}
\newcommand{\oelem}[1]{{\color[RGB]{0,0,200}\uwave{#1}}}

 \begin{table}[]
	\small
	\centering
	\caption{For three samples of the test set (S), we show the most similar dynamic example (D). Gold labels are highlighted for \pelem{Participants}, \ielem{Interventions}, and \oelem{Outcomes} in all sentences. Note that to workers only the labels for either P, I, or O (depending on the sub-task) within the dynamic examples are visible.}
	\label{tab:examples}
	\beforetable{}
	\begin{tabularx}{\linewidth}{@{}X@{}}
		\textbf{S:} We performed a randomized, controlled study comparing the \oelem{prophylactic effects} of capsule forms of \ielem{fluconazole} (n=110) and \ielem{itraconazole} (n=108) in \pelem{patients with acute myeloid leukemia (AML) or myelodysplastic syndromes (MDS) during and after chemotherapy}. \\  \addlinespace[3pt]
		\textbf{D:} A randomized, double-blind, placebo-controlled study on the \oelem{immediate clinical and microbiological efficacy} of \ielem{doxycycline} (100mg for 14 days) was carried out to determine the benefit of adjunctive medication in \pelem{16 patients with localized juvenile periodontitis}. \\ \midrule[1.5pt]
		\textbf{S:} \oelem{Adverse events} did not significantly differ in the 2 groups. \\ \addlinespace[3pt]
		\textbf{D:} There were no serious \oelem{adverse events}. \\ \midrule[1.5pt]
		\textbf{S:} The majority (63\%) of the project group had no \oelem{admission} during the 10 month study period. \\ \addlinespace[3pt]
		\textbf{D:} Referral occurred at any stage of the patients' EECU admission. \\   
	\end{tabularx}
	\aftertable{}
\end{table}

In Table~\ref{tab:examples}, we illustrate three sample sentences of the test set and the corresponding most similar dynamic example of the training set. The first case shows that the dynamic example provides strong support in annotating P, I and, O -- even though the sentence is rather complex and long. The middle case shows a dynamic example that provides support in annotating the O element. Finally, the last case shows that no appropriate dynamic example is found for the sample. In such cases, workers need to decide independently.

\subsection{\tra{} Approach}
We compare annotations obtained via the \zla{} approach to the non-expert annotations that were previously obtained in the scope of~\cite{nyeCorpusMultiLevelAnnotations2018} using the \tra{} approach\footnote{Annotations of individual crowdworkers were downloaded from \url{https://github.com/bepnye/EBM-NLP}}. Note that we decided to re-use the available annotations rather than re-collecting them, since the same annotation setup is followed in both approaches (Section \ref{sec:piosetup}). Although a different annotation interface was considered by~\cite{nyeCorpusMultiLevelAnnotations2018}, the interaction component of clicking a start and end word to annotate a P, I, or O text phrase is identical in both approaches.

\sectionspace{}
\section{Results}
\label{sec:results}
We compare the \zla{} approach to the \tra{} approach based on the 426 sentences of the test set. For the \tra{} approach, at least 3 redundant non-expert annotations are available per sentence; although, the average is $\approx 11$. For the \zla{} approach, we collect exactly 3 redundant annotations per sentence, resulting in a total of $426 \times 3=1,278$  sentence annotations per PIO sub-task. In total, 26 workers contributed in annotating the test set using the \zla{} approach. In contrast, 403 workers contributed for the \tra{} approach ~\cite{nyeCorpusMultiLevelAnnotations2018}, because of (i) the goal of collecting more redundant samples and (ii) an additional goal of labelling $\approx 5,000$ clinical trials only by non-experts.

\subsubsection*{Agreement of Individuals}
We compare annotations obtained by \zla{} and \tra{} based on the inter-annotator agreement to the gold standard annotations using Cohen's Kappa. To eliminate random noise of workers who labeled only a few sentences, we do not analyze workers of \zla{} and \tra{} who labeled less than 5\% of the total of 426 test sentences ($\approx 21$ sentences). 

\begin{figure}[h!]
	\centering
	\includegraphics[width=\linewidth]{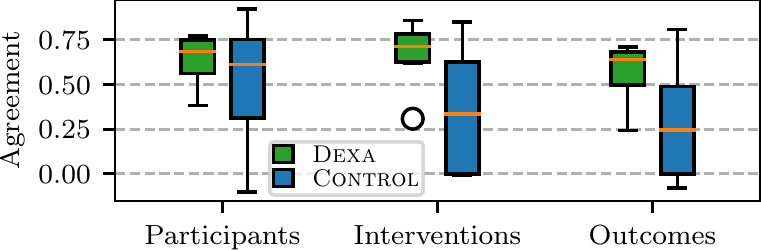}
	\caption{Cohen's $\kappa$ between annotations of individual non-expert annotators compared to the gold standard.}
	\label{fig:boxplot}
\end{figure}

The results in Figure~\ref{fig:boxplot} show that the median agreement to experts is substantially higher in the \zla{} approach than in the \tra{} approach, especially for I and O. Notice that the kappa scores $\kappa$ of workers of the \tra{} approach range from 0.0 to nearly 1.0, probably affected by the higher number of redundantly collected labels. Notable is also the one worker of the \zla{} approach who underperformed in the I sub-task compared to the other workers, illustrated as a dot.

\vspace{-4pt}
\vspace{-4pt}

\subsubsection*{Agreement of Aggregations}
We analyze quality of aggregated annotations obtained with two aggregation strategies: majority voting (MV), which weights individual workers equally, and the Dawid-Skene (DS) model~\cite{dawidMaximumLikelihoodEstimation1979}, which recognizes reliable annotators and weights them more strongly in the aggregation procedure. For \zla{}, we  aggregate the 3 available labels, while for \tra{}, we create aggregations from different numbers of labels $n=\{3,6,9, ALL\}$, as follows: (i) for each task-instance, we randomly pick $n$ annotations from the available $~\approx11$ redundant ones; (ii) we repeat step \textit{(i)} 20 times and compute the agreement to the gold standard label at each iteration; (iii) the 20 $\kappa$-values are averaged to compute the final $\kappa$. 

The results in Table~\ref{tab:mjr} show that the $\kappa$ score between non-experts and experts is in almost all cases higher for aggregated annotations obtained with \zla{} compared to \tra{}. Even when 6, 9 or all  labels of the \tra{} approach are aggregated, the 3 aggregated annotations of the \zla{} approach reach a higher agreement to the gold standard for the I and O sub-tasks. Only for P, we observed that the aggregation of all redundant \tra{} annotations surpasses the $\kappa$ score of our \zla{} approach.

Notable is the effectiveness of the DS aggregation for the redundant labels of the \tra{} approach. Especially for P, the high agreements of individual workers using the \tra{} approach (Figure~\ref{fig:boxplot}) leads to a strong aggregated result via DS (Table~\ref{tab:mjr}). No improvements are observed when using DS over MV to aggregate the 3 redundant \zla{} labels, which is expected since the noise of individual annotators is low (as shown in Figure~\ref{fig:boxplot}).

\vspace{-4pt}
\begin{table}[h!]
	\centering
		\caption{Cohen's $\kappa$ between the gold standard and non-expert annotations aggregated for different numbers of annotations $n$ via Majority Vote (MV) and Dawid-Skene (DS).}
	\label{tab:mjr}
	\beforetable{}
	\begin{tabular}{lrrr}
		\toprule
		& \multicolumn{3}{c}{Cohen's Kappa ($\kappa$)} \\ \cmidrule(lr){2-4}
		&    P &    I &    O \\ \midrule
\zla{}$_{MV3}$      &          \textbf{0.780} &           \textbf{0.757} &      \textbf{0.694} \\
\crowd{}$_{MV3}$    &          0.702 &           0.455 &      0.352 \\
\crowd{}$_{MV6}$    &          0.729 &           0.465 &      0.342 \\
\crowd{}$_{MV9}$    &          0.749 &           0.454 &      0.307 \\
\crowd{}$_{MV ALL}$ &          0.746 &           0.457 &      0.311 \\ \midrule
\zla{}$_{DS3}$      &          0.776 &           \textbf{0.756} &      \textbf{0.694} \\
\crowd{}$_{DS3}$    &          0.729 &           0.579 &      0.458 \\
\crowd{}$_{DS6}$    &          0.809 &           0.644 &      0.614 \\
\crowd{}$_{DS9}$    &          0.841 &           0.629 &      0.659 \\
\crowd{}$_{DS ALL}$ &         \textbf{0.867} &           0.633 &      0.677 \\

		\bottomrule
	\end{tabular}
	\aftertable{}
\end{table}

\subsubsection*{Worker Feedback} We analyze the feedback from workers on the usefulness of the dynamic examples in Table~\ref{tab:useful}. The result show a high percentage of positive answers for all annotation tasks, especially for the more difficult tasks I (78\%) and O (76\%). Additionally, the (perceived) usefulness of the examples has an effect on the quality of the annotations. Indeed, the averaged agreements of individual annotators (excluding workers who annotated less than 5\% of the 426 test sentences) to the gold standard labels is on average much higher when the dynamic examples were found useful than otherwise. 

\begin{table}[h!]
	\centering
	\caption{Percentage of workers finding dynamic examples useful; average $\kappa$ scores (std. deviation) to the gold standard.}
	\label{tab:useful}
	\small
	\beforetable{}
	\begin{tabular}{lrrrrrr}
		\toprule
		\multirow{2}{*}{Feedback} & \multicolumn{3}{c}{Percentage} & \multicolumn{3}{c}{Cohen's Kappa ($\kappa$)} \\ \cmidrule(lr){2-4} \cmidrule(lr){5-7}
		& P &    I &    O&    P &    I &    O \\ \midrule
		Useful    & 64\%  &            78\%  &       76\% & 0.73$\pm.12$  &            0.67$\pm.14$  &       0.60$\pm.18$  \\
		Not useful    & 36\%  &            22\%  &       24\% &           0.42$\pm.07$  &            0.41$\pm.14$  &      0.44$\pm.18$  \\
		\bottomrule
	\end{tabular}
	\aftertable{}
\end{table}

\sectionspace{}
\section{Conclusion}
\label{sec:concl}
We presented the \zla{} annotation approach in which non-expert annotators are supported not only by task level annotation examples (as in \tra{}) but also by dynamic, task-instance level examples that are semantically similar to the currently annotated sample. Evaluating \zla{} on the PIO task lead to: (i) \textit{improved quality of individual annotations}: individual annotator agreement with expert annotations was on average $0.28$ higher for \zla{} than \tra{}; (ii) \textit{improved aggregated label quality}: three per majority voting aggregated annotations of the \zla{} approach reached on average $0.15$ higher agreements to experts than in the \tra{} approach; (iii) \textit{explicit validation of dynamic example usefulness}: workers found the proposed examples useful in the majority of cases (avg. 73\% over PIO tasks) and label quality was consistently higher for cases when the examples were judged useful than otherwise.

As future work we will (i) optimize the parameter $k$; (ii) investigate the effectiveness of different similarity methods for selecting examples through A/B testing, and (iii) evaluate \zla{} on different domains and annotation tasks.
\vspace{-8pt}

\bibliographystyle{ACM-Reference-Format}
\bibliography{bib}

\end{document}